\documentclass[journal,onecolumn, 11pt]{IEEEtran}

\usepackage{graphicx}

\usepackage{amsmath}

\newcommand{\aasd}{{a_{sd}}}
\newcommand{\aard}{{a_{rd}}}
\newcommand{\aasr}{{a_{sr}}}
\newcommand{\asd}{{|a_{sd}|}}
\newcommand{\ard}{{|a_{rd}|}}
\newcommand{\asr}{{|a_{sr}|}}
\newcommand{\cN}{{\mathcal N}}
\newcommand{\E}{{\cal E}}

\newcommand{\dotleq}{{\stackrel{.}{\leq}}}

\newtheorem{theorem}{Theorem}
\newtheorem{lemma}{Lemma}
\newtheorem{myrule}{Policy}
\newtheorem{definition}{Definition}

\begin{document}
\title{A Simple Cooperative Diversity Method Based on Network Path Selection}
%
%

\author{Aggelos~Bletsas,
        ~Ashish~Khisti,
        ~David~P.~Reed,
        ~Andrew Lippman\\
        Massachusetts Institute of Technology \\
        aggelos@media.mit.edu~~~khisti@mit.edu \\
        \thanks{This work was supported by NSF under grant number CNS-0434816,
        the MIT Media Laboratory Digital Life Program and a Nortel Networks
graduate fellowship award. Submitted on January 2005 to IEEE
Journal Selected Areas of Communication, special issue on 4G,
revised April 2005, accepted July 2005. Expected publication
second quarter of 2006.}}



%



\maketitle

\begin{abstract}
Cooperative diversity has been recently proposed as a way to form
virtual antenna arrays that provide dramatic gains in slow fading
wireless environments. However most of the proposed solutions
require distributed space-time coding algorithms, the careful
design of which is left for future investigation if there is more
than one cooperative relay. We propose a novel scheme, that
alleviates these problems and provides diversity gains on the
order of the number of relays in the network. Our scheme first
selects the best relay from a set of $M$ available relays and then
uses this "best" relay for cooperation between the source and the
destination. We develop and analyze a distributed method to select
the best relay that requires no topology information and is based
on local measurements of the instantaneous channel conditions.
This method also requires no explicit communication among the
relays. The success (or failure) to select the best available path
depends on the statistics of the wireless channel, and a
methodology to evaluate performance for any kind of wireless
channel statistics, is provided. Information theoretic analysis of
outage probability shows that our scheme achieves the same
diversity-multiplexing tradeoff as achieved by more complex
protocols, where coordination and distributed space-time coding
for $M$ nodes is required, such as those proposed in
\cite{LanemanWornell03}. The simplicity of the technique, allows
for immediate implementation in existing radio hardware and its
adoption could provide for improved flexibility, reliability and
efficiency in future 4G wireless systems.
\end{abstract}


%

\pagestyle{plain}

\section{Introduction}
\PARstart{I}{n} this work, we propose and analyze a practical
scheme that forms a virtual antenna array among single antenna
terminals, distributed in space. The setup includes a set of
cooperating relays which are willing to forward received
information towards the destination and the proposed method is
about a distributed algorithm that selects the most appropriate
relay to forward information towards the receiver. The decision is
based on the end-to-end instantaneous wireless channel conditions
and the algorithm is distributed among the cooperating wireless
terminals.

The best relay selection algorithm lends itself naturally into
cooperative diversity protocols
\cite{SendonarisErkipAazhang03-1,SendonarisErkipAazhang03-2,LanemanTseWornell04,Hunter-Nosratina},
which have been recently proposed to improve reliability in
wireless communication systems using distributed virtual antennas.
The key idea behind these protocols is to create additional paths
between the source and destination using intermediate relay nodes.
In particular, Sendonaris, Erkip and Aazhang
\cite{SendonarisErkipAazhang03-1},
\cite{SendonarisErkipAazhang03-2} proposed a way of
\textit{beamforming} where source and a cooperating relay,
assuming knowledge of the forward channel, adjust the phase of
their transmissions so that the two copies can add coherently at
the destination. Beamforming requires considerable modifications
to existing RF front ends that increase complexity and cost.
Laneman, Tse and Wornell \cite{LanemanTseWornell04} assumed no CSI
at the transmitters and therefore assumed no beamforming
capabilities and proposed the analysis of cooperative diversity
protocols under the framework of diversity-multiplexing tradeoffs.
Their basic setup included one sender, one receiver and one
intermediate relay node and both analog as well as digital
processing at the relay node were considered. Subsequently, the
diversity-multiplexing tradeoff of cooperative diversity protocols
with multiple relays was studied in
\cite{LanemanWornell03,AzarianElGamal04}. While
\cite{LanemanWornell03} considered the case of orthogonal
transmission\footnote{Note that in that scheme the relays do not
transmit in mutually orthogonal time/frequency bands. Instead they
use a space-time code to collaboratively send the message to the
destination. Orthogonality refers to the fact that the source
transmits in time slots orthogonal to the relays. Throughout this
paper we will refer to Laneman's scheme as orthogonal cooperative
diversity. } between source and relays, \cite{AzarianElGamal04}
considered the case where source and relays could transmit
simultaneously. It was shown in \cite{AzarianElGamal04} that by
relaxing the orthogonality constraint, a considerable improvement
in performance could be achieved, albeit at a higher complexity at
the decoder. These approaches were however information theoretic
in nature and the design of practical codes that approach these
limits was left for further investigation.

Such code design is difficult in practice and an open area of
research: while space time codes for the Multiple Input Multiple
Output (MIMO) link do exist \cite{GamalCaireDamen} (where the
antennas belong to the same central terminal), more work is needed
to use such algorithms in the relay channel, where antennas belong
to different terminals distributed in space. The relay channel is
fundamentally different than the point-to-point MIMO link since
information is not \textit{a priori} known to the cooperating
relays but rather needs to be communicated over noisy links.
Moreover, the number of participating antennas is not fixed since
it depends on how many relay terminals participate and how many of
them are indeed useful in relaying the information transmitted
from the source. For example, for relays that decode and forward,
it is necessary to decode successfully before retransmitting. For
relays that amplify and forward, it is important to have a good
received SNR, otherwise they would forward mostly their own noise
\cite{Nabar}. Therefore, the number of participating antennas in
cooperative diversity schemes is in general random and space-time
coding invented for fixed number of antennas should be
appropriately modified. It can be argued that for the case of
orthogonal transmission studied in the present work (i.e.
transmission during orthogonal time or frequency channels) codes
can be found that maintain orthogonality in the absence of a
number of antennas (relays). That was pointed in
\cite{LanemanWornell03} where it was also emphasized that it
remains to be seen how such codes could provide residual diversity
without sacrifice of the achievable rates. In short, providing for
practical space-time codes for the cooperative relay channel is
fundamentally different than space-time coding for the MIMO link
channel and is still an open and challenging area of research.

Apart from practical space-time coding for the cooperative relay
channel, the formation of virtual antenna arrays using individual
terminals distributed in space, requires significant amount of
coordination. Specifically, the formation of cooperating groups of
terminals involves distributed algorithms \cite{LanemanWornell03}
while synchronization at the packet level is required among
several different transmitters. Those additional requirements for
cooperative diversity demand significant modifications to almost
all layers of the communication stack (up to the routing layer)
which has been built according to "traditional", point-to-point
(non-cooperative) communication.

\begin{figure}
  \centering
  \includegraphics[width=3.0in]{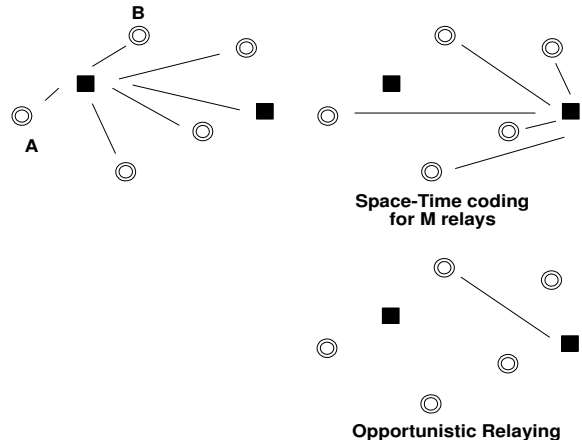}\\
  \caption{A transmission is overheard by neighboring nodes. Distributed Space-Time
  coding is needed so that all overhearing nodes could simultaneously
  transmit. In this work we analyze "Opportunistic Relaying" where the relay
  with the strongest transmitter-relay-receiver path is selected,
  among several candidates, in a distributed fashion using
  instantaneous channel measurements.}
  \label{fig:motivation}
\end{figure}

In fig. \ref{fig:motivation} a transmitter transmits its
information towards the receiver while all the neighboring nodes
are in listening mode. For a practical cooperative diversity in a
three-node setup, the transmitter should know that allowing a
relay at location (B) to relay information, would be more
efficient than repetition from the transmitter itself. This is not
a trivial task and such event depends on the wireless channel
conditions between transmitter and receiver as well as between
transmitter-relay and relay-receiver. What if the relay is located
in position (A)? This problem also manifests in the multiple relay
case, when one attempts to simplify the physical layer protocol by
choosing the best available relay. In \cite{Valenti} it was
suggested that the best relay be selected based on location
information with respect to source and destination based on ideas
from geographical routing proposed in \cite{Zorzi:a}. Such schemes
require knowledge or estimation of distances between all relays
and destination and therefore require either a) infrastructure for
distance estimation (for example GPS receivers at each terminal)
or b) distance estimation using expected SNRs which is itself a
non-trivial problem and is more appropriate for static networks
and less appropriate for mobile networks, since in the latter
case, estimation should be repeated with substantial overhead.

In contrast, we propose a novel scheme that selects the best relay
between source and destination based on {\em instantaneous}
channel measurements. The proposed scheme requires no knowledge of
the topology or its estimation. The technique is based on signal
strength measurements rather than distance and requires a small
fraction of the channel {\it coherence time}. All these features
make the design of such a scheme highly challenging and the
proposed solution becomes non-trivial. Additionally, the algorithm
itself provides for the necessary coordination in time and group
formation among the cooperating terminals.


The three-node reduction of the multiple relay problem we
consider, greatly simplifies the physical layer design. In
particular, the requirement of space-time codes is completely
eliminated if the source and relay transmit in orthogonal
time-slots. We further show that there is essentially no loss in
performance in terms of the diversity-multiplexing tradeoff as
compared to the  transmission scheme in \cite{LanemanWornell03}
which requires space-time coding across the relays successful in
decoding the source message. We also note that our scheme can be
used to simplify the non-orthogonal multiple relay protocols
studied in \cite{AzarianElGamal04}. Intuitively, the gains in
cooperative diversity do not come from using complex schemes, but
rather from the fact that we have enough relays in the system to
provide sufficient diversity. 

The simplicity of the technique, allows for immediate
implementation in existing radio hardware. An implementation of
the scheme using custom radio hardware is reported in
\cite{Bletsas05_thesis}. Its adoption could provide for improved
flexibility (since the technique addresses coordination issues),
reliability and efficiency (since the technique inherently builds
upon diversity) in future 4G wireless systems.

\subsection{Key Contributions}
 One of the key contribution of this paper is to propose and
 analyze a simplification of user cooperation protocols at the
 physical layer by using a smart relay selection algorithm at the
 network layer. Towards this end, we take the following steps:
 \begin{itemize}
 \item A new protocol for selection of the "best" relay between
 the source and destination is suggested and analyzed. This
 protocol has the following features:
 \begin{itemize}
    \item The protocol is distributed and each relay only makes local
    channel measurements.

    \item Relay selection is based on instantaneous channel
    conditions in slow fading wireless environments.
    No prior knowledge of topology or estimation of it is
    required.

    \item The amount of overhead involved in selecting the best relay is minimal.
    It is shown that there is a flexible tradeoff between the time incurred in the protocol
    and the resulting error probability.
 \end{itemize}

\item The impact of smart relaying on the performance of user
cooperation protocols is studied. In particular, it is shown that
for orthogonal cooperative diversity protocols there is no loss in
performance (in terms of the diversity multiplexing tradeoff) if
only the best relay participates in cooperation. Opportunistic
relaying provides an alternative solution with a very simple
physical layer to conventional cooperative diversity protocols
that rely on space-time codes. The scheme could be further used to
simplify space-time coding in the case of non-orthogonal
transmissions.
\end{itemize}

Since the communication scheme exploits the wireless channel at
its best, via distributed cooperating relays,  we naturally called
it \textit{opportunistic relaying}. The term "opportunistic" has
been widely used in various different contexts. In \cite{Biswas},
it was used in the context of repetitive transmission of the same
information over several paths, in 802.11b networks. In our setup,
we do not allow repetition since we are interested in providing
diversity without sacrificing the achievable rates, which is a
characteristic of repetition schemes. The term "opportunistic" has
also been used in the context of efficient \textit{flooding} of
signals in multi-hop networks \cite{Scaglione}, to increase
communication range and therefore has no relationship with our
work. We first encountered the term "opportunistic" in the work by
Viswanath, Tse and Laroia \cite{ViswanathTseLaroia}, where the
base station always selects the best user for transmission in an
artificially induced fast fading environment.  In our work, a
mechanism of multi-user diversity is provided for the relay
channel, in single antenna terminals. Our proposed scheme,
resembles \textit{selection diversity} that has been proposed for
centralized multi-antenna receivers \cite{Molisch04}. In our
setup, the single antenna relays are distributed in space and
attention has been given in selecting the ``best" possible
antenna, well before the channel changes again, using minimal
communication overhead.

In the following section, we describe in detail opportunistic
relaying and contrast its distributed, location information-free
nature to existing approaches in the field. Probabilistic analysis
and close form expressions regarding the success (or failure) and
speed of ``best" path selection, for any kind of wireless channel
statistics, are provided in section \ref{Probabilistic Analysis}.
In section \ref{Simplifying} we prove that opportunistic relaying
has no performance loss compared to complex space-time coding,
under the same assumptions of orthogonal channel transmissions
\cite{LanemanWornell03} and discuss the ability of the scheme to
further simplify space-time coding for non-orthogonal channels. We
also discuss in more detail, why space-time codes designed for the
MIMO link are not directly applicable to the cooperative relay
channel. We conclude in section \ref{conclusion}.

\section{Description of Opportunistic Relaying}
\label{description}
\begin{figure}
\centering
  \includegraphics[width=4.0in]{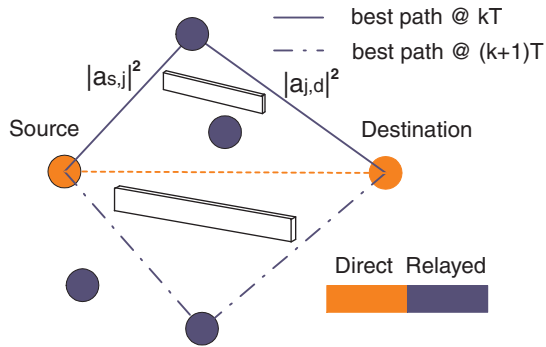}\\
  \caption{Source transmits to destination and neighboring nodes
  overhear the communication. The "best" relay among M candidates is selected to
  relay information, via a distributed mechanism and based on instantaneous
  end-to-end channel conditions. For the diversity-multiplexing tradeoff
  analysis, transmission of source and "best" relay occur in
  orthogonal time channels. The scheme could be easily modified to incorporate
  simultaneous transmissions from source and "best" relay.}\label{fig:setup}
\end{figure}
According to opportunistic relaying, a single relay among a set of
$M$ relay nodes is selected, depending on which relay provides for
the "best" end-to-end path between source and destination (fig.
\ref{fig:motivation}, \ref{fig:setup}). The wireless channel
$a_{si}$ between source and each relay $i$, as well as the channel
$a_{id}$ between relay $i$ and destination affect performance.
These parameters model the propagation environment between any
communicating terminals and change over time, with a rate that
macroscopically can be modelled as the \textit{Doppler shift},
inversely proportional to the channel \textit{coherence time}.
Opportunistic selection of the ``best" available relay involves
the discovery of the most appropriate relay, in a distributed and
``quick" fashion, well before the channel changes again. We will
explicitly quantify the speed of relay selection in the following
section.

The important point to make here is that under the proposed
scheme, the relay nodes monitor the \textit{instantaneous} channel
conditions towards source and destination, and decide in a
distributed fashion which one has the \textit{strongest} path for
information relaying, well before the channel changes again. In
that way, topology information at the relays (specifically
location coordinates of source and destination at each relay) is
not needed. The selection process \textit{reacts} to the physics
of wireless propagation, which are in general dependent on several
parameters including mobility and distance. By having the network
select the relay with the strongest end-to-end path,
\textit{macroscopic} features like ``distance" are also taken into
account. Moreover, the proposed technique is advantageous
 over techniques that select the best relay \textit{a priori},
 based on distance toward source or destination, since distance-dependent relay selection neglects
  well-understood phenomena in wireless propagation such as
\textit{shadowing} or \textit{fading}: communicating
transmitter-receiver pairs with similar distances might have
enormous differences in terms of received SNRs.  Furthermore,
average channel conditions might be less appropriate for mobile
terminals than static. Selecting the best available path under
such conditions (zero topology information, "fast" relay selection
well bellow the coherence time of the channel and minimum
communication overhead) becomes non-obvious and it is one of the
main contributions of this work.

More specifically, the relays overhear a single transmission of a
Ready-to-Send (RTS) packet and a Clear-to-Send (CTS) packet from
the destination. From these packets the relays assess how
appropriate each of them is for information relaying. The
transmission of RTS from the source allows for the estimation of
the instantaneous wireless channel $a_{si}$ between source and
relay $i$, at each relay $i$ (fig. \ref{fig:setup}). Similarly,
the transmission of CTS from the destination, allows for the
estimation of the instantaneous wireless channel $a_{id}$ between
relay $i$ and destination, at each relay $i$, according to the
reciprocity theorem\cite{Rappaport96}\footnote{We assume that the
forward and backward channels between the relay and destination
are the same from the reciprocity theorem. Note that these
transmissions occur on the same frequency band and same coherence
interval.}. Note that the source does not need to listen to the
CTS packet\footnote{The CTS packet name is motivated by   existing
MAC protocols. However unlike the existing MAC protocols,the
source does not need to receive this packet.} from the
destination.


Since communication among all relays should be minimized for
reduced overall overhead, a method based on time was selected: as
soon as each relay receives the CTS packet, it starts a timer from
a parameter $h_i$ based on the instantaneous channel measurements
$a_{si},a_{id}$. The timer of the relay with the best end-to-end
channel conditions will expire first. That relay transmits a short
duration \textit{flag} packet, signaling its presence. All relays,
while waiting for their timer to reduce to zero (i.e. to expire)
are in listening mode. As soon as they hear another relay to flag
its presence or forward information (the best relay), they back
off.

For the case where all relays can listen source and destination,
but they are "hidden" from each other (i.e. they can not listen
each other), the best relay notifies the destination with a short
duration \textit{flag} packet and the destination notifies all
relays with a short broadcast message.

The channel estimates $a_{si}, ~a_{id}$ at each relay, describe
the quality of the wireless path between source-relay-destination,
for each relay $i$. Since the two hops are both important for
end-to-end performance, each relay should quantify its
appropriateness as an active relay, using a function that involves
the link quality of both hops. Two functions are used in this
work: under policy I, the minimum of the two is selected (equation
(\ref{eq:policy1})), while under policy II, the harmonic mean of
the two is used (equation (\ref{eq:policy2})). Policy I selects
the "bottleneck" of the two paths while Policy II balances the two
link strengths and it is a smoother version of the first one.

Under policy I:

\vspace{-2em}

\begin{equation}\label{eq:policy1}
    h_i=\min\{|a_{si}|^2,|a_{id}|^2\}
\end{equation}

Under policy II:

\vspace{-2em}

\begin{equation}\label{eq:policy2}
    h_i=\frac{2}{\frac{1}{|a_{si}|^2} + \frac{1}{|a_{id}|^2}}=\frac{2~|a_{si}|^2 ~ |a_{id}|^2}{|a_{si}|^2+|a_{id}|^2}
\end{equation}

The relay $i$ that maximizes function $h_i$ is the one with the
"best" end-to-end path between initial source and final
destination. After receiving the CTS packet, each relay $i$ will
start its own timer with an initial value $T_i$, inversely
proportional to the end-to-end channel quality $h_i$, according to
the following equation:

\vspace{-2em}

\begin{equation}\label{eq:timer}
    T_i=\frac{\lambda}{h_i}
\end{equation}
Here $\lambda$ is a constant. The units of $\lambda$ depend on the
units of $h_i$.  Since $h_i$ is a scalar, $\lambda$ has the units
of time. For the discussion in this work, $\lambda$ has simply
values of $\mu secs$.
\begin{eqnarray}\label{eq:h_b}
  h_b &=& \max\{h_i\}, ~\Longleftrightarrow\\ \label{eq:h_b2}
  T_b &=& \min\{T_i\}, ~i~\in~[1..M].
\end{eqnarray}

Therefore, the "best" relay has its timer reduced to zero first
(since it started from a smaller initial value, according to
equations (\ref{eq:timer})-(\ref{eq:h_b2}). This is the relay $b$
that participates in forwarding information from the source. The
rest of the relays, will overhear the "flag" packet from the best
relay (or the destination, in the case of hidden relays) and back
off.

After the best relay has been selected, then it can be used to
forward information towards the destination. Whether that "best"
relay will transmit simultaneously with the source or not, is
completely irrelevant to the relay selection process. However, in
the diversity-multiplexing tradeoff analysis in section
\ref{Simplifying}, we strictly allow only one transmission at each
time and therefore we can view the overall scheme as a two-step
transmission: one from source and one from "best" relay, during a
subsequent (orthogonal) time channel (fig. \ref{fig:setup}).

\subsection{A note on Time Synchronization}
In principle, the RTS/CTS transmissions between source and
destination, existent in many Medium Access Control (MAC)
protocols, is only needed so that all intermediate relays can
assess their connectivity paths towards source and destination.
The reception of the CTS packet triggers at each relay the
initiation of the timing process, within an uncertainty interval
that depends on different propagation times, identified in detail
in section \ref{Probabilistic Analysis}. Therefore, an explicit
time synchronization protocol among the relays is not required.
Explicit time synchronization would be needed between source and
destination, only if there was no direct link between them. In
that case, the destination could not respond with a CTS to a RTS
packet from the source,  and therefore source and destination
would need to \textit{schedule} their RTS/CTS exchange by other
means.  In such cases "crude" time synchronization would be
useful. Accurate synchronization schemes, centralized
\cite{Bletsas03} or decentralized \cite{Bletsas05}, do exist and
have been studied elsewhere. We will assume that source and
destination are in communication range and therefore no
synchronization protocols are needed.

\subsection{A note on Channel State Information (CSI)}
CSI at the relays, in the form of link strengths (not signal
phases), is used at the network layer for "best" relay selection.
CSI is not required at the physical layer and is not exploited
either at the source or the relays. The wireless terminals in this
work do not exploit CSI for \textit{beamforming} and do not adapt
their transmission rate to the wireless channel conditions, either
because they are operating in the minimum possible rate or because
their hardware do not allow multiple rates. We will emphasize
again that no CSI at the physical layer is exploited at the source
or the relays, during the diversity-multiplexing tradeoff
analysis, in section \ref {Simplifying}.

\subsection{Comparison with geometric approaches}
As can be seen from the above equations, the scheme depends on the
instantaneous channel realizations or equivalently, on received
\textit{instantaneous} SNRs, at each relay. An alternative
approach would be to have the source know the location of the
destination and propagate that information, alongside with its own
location information to the relays, using a simple packet that
contained that location information. Then, each relay, assuming
knowledge of its own location information, could assess its
proximity towards source and destination and based on that
proximity, contend for the channel with the rest of the relays.
That is an idea, proposed by Zorzi and Rao \cite{Zorzi:a} in the
context of fading-free wireless networks, when nodes know their
location and the location of their destination (for example they
are equipped with GPS receivers). The objective there was to study
geographical routing and study the average number of hops needed
under such schemes. All relays are partitioned into a specific
number of geographical regions between source and destination and
each relay identifies its region using knowledge of its location
and the location of source and destination. Relays at the region
closer to the destination contend for the channel first using a
standard CSMA splitting scheme. If no relays are found, then
relays at the second closest region contend and so on, until all
regions are covered, with a typical number of regions close to 4.
The latency of the above distance-dependent contention resolution
scheme was analyzed in \cite{Zorzi:b}.

Zorzi and Rao's scheme of distance-dependent relay selection was
employed in the context of Hybrid-ARQ, proposed by Zhao and
Valenti \cite{Valenti}. In that work, the request to an Automatic
Repeat Request (ARQ) is served by the relay closest to the
destination, among those that have decoded the message. In that
case, code combining is assumed that exploits the direct and
relayed transmission (that's why the term \textit{Hybrid} was
used)\footnote{The idea of having a relay terminal respond to an
ARQ instead of the original source, was also reported and analyzed
in \cite{LanemanTseWornell04} albeit for repetition coding instead
of hybrid code combining.}. Relays are assumed to know their
distances to the destination (valid for GPS equipped terminals) or
estimate their distances by measuring the expected channel
conditions using the ARQ requests from the destination or using
other means.

We note that our scheme of opportunistic relaying differs from the
above scheme in the following aspects:

\begin{itemize}
\item The above scheme performs relay selection based on
geographical regions while our scheme performs selection based on
instantaneous channel conditions. In wireless environment, the
latter choice could be more suitable as relay nodes located at
similar distance to the destination could have vastly different
channel gains due to effects such as fading.

\item The above scheme requires measurements to be only performed
once if there is no mobility among nodes but requires several
rounds of packet exchanges to determine the average SNR. On the
other hand opportunistic relaying requires only three packet
exchanges in total to determine the instantaneous SNR, but
requires that these measurements be repeated in each coherence
interval. We show in the subsequent section that the overhead of
relay selection is a small fraction of the coherence interval with
collision probability less than $0.6 \%$.

\item We also  note that our protocol is a proactive protocol
since it selects the best relay before transmission. The protocol
can easily be made to be reactive (similar to \cite{Valenti}) by
selecting the relay after the first phase. However this
modification would require all relays to listen to the source
transmission which can be energy inefficient from a network sense.

\end{itemize}

%
\section{Probabilistic Analysis of Opportunistic Relaying}
\label{Probabilistic Analysis}
The probability of having two or
more relay timers expire "at the same time" is zero. However, the
probability of having two or more relay timers expire within the
same time interval $c$ is non zero and can be analytically
evaluated, given knowledge of the wireless channel statistics.

The only case where opportunistic relay selection fails is when
one relay can not detect that another relay is more appropriate
for information forwarding. Note that we have already assumed that
all relays can listen initial source and destination, otherwise
they do not participate in the scheme. We will assume two extreme
cases: a) all relays can listen to each other b) all relays are
hidden from each other (but they can listen source and
destination). In that case, the flag packet sent by the best relay
is received from the destination which responds with a short
broadcast packet to all relays. Alternatively, other schemes based
on "busy tone" (secondary frequency) control channels could be
used, requiring no broadcast packet from the destination and
partly alleviating the "hidden" relays problem.

\begin{figure}
\centering
\includegraphics[width=3.0in]{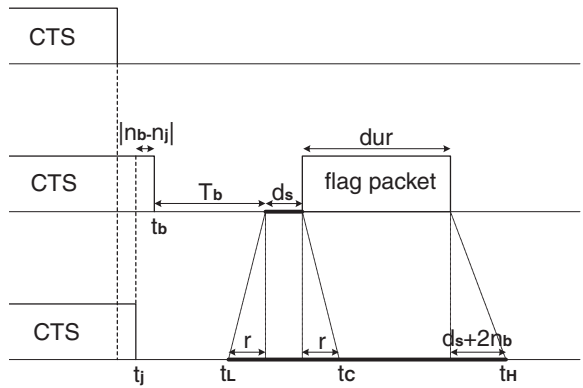}
\caption{The middle row corresponds to the "best" relay. Other
relays (top or bottom row) could erroneously be selected as "best"
relays, if their timer expired within intervals when they can not
hear the best relay transmission. That can happen in the interval
$[t_L, t_C]$ for case (a) (No Hidden Relays) or $[t_L, t_H]$ for
case (b) (Hidden Relays). $t_b, t_j$ are time points where
reception of the CTS packet is completed at best relay $b$ and
relay $j$ respectively.} \label{fig_mac}
\end{figure}

From fig. \ref{fig_mac}, collision of two or more relays can
happen if the best relay timer $T_b$ and one or more other relay
timers expire within $[t_L, ~t_C]$ for the case of no hidden
relays (case (a)). This interval depends on the radio switch time
from receive to transmit mode $d_s$ and the propagation times
needed for signals to travel in the wireless medium. In custom
low-cost transceiver hardware, this switch time is typically on
the order of a few $\mu seconds$ while propagation times for a
range of 100 meters is on the order of $1/3~\mu seconds$. For the
case of "hidden" relays the uncertainty interval becomes $[t_L,
~t_H]$ since now the duration of the flag packet should be taken
into account, as well as the propagation time towards the
destination and back towards the relays and the radio switch time
at the destination. The duration of the flag packet can be made
small, even one bit transmission could suffice. In any case, the
higher this uncertainty interval, the higher the probability of
two or more relay timers to expire within that interval. That's
why we will assume maximum values of $c$, so that we can assess
worst case scenario performance.

(a) No Hidden Relays:
\begin{equation}\label{eq:c_nohidden}
    c=r_{max}+|n_b-n_j|_{max}+d_s
\end{equation}

(b) Hidden Relays:
\begin{equation}\label{eq:c_hidden}
    c=r_{max}+|n_b-n_j|_{max}+2d_s + dur + 2n_{max}
\end{equation}
\begin{itemize}
    \item $n_j$: propagation delay between relay $j$ and
    destination. $n_{max}$ is the maximum.
    \item $r$: propagation delay between two relays. $r_{max}$ is
    the maximum.
    \item $d_s$: receive-to-transmit switch time of each radio.
    \item $dur$: duration of flag packet, transmitted by "best"
    relay.
\end{itemize}
In any case, the probability of having two or more relays expire
within the same interval $c$, out of a collection of $M$ relays,
can be described by the following expression:
\begin{eqnarray}\label{eq:upper_bound}
    Pr(Collision) & \leq & Pr(any ~T_j < T_b+c~|~j \neq b) \\
    \nonumber
    where~T_b &=& \min\{T_j\},~j \in [1,M] ~and~ c>0.
\end{eqnarray}

Notice that we assume failure of relay selection when two or more
relays collide. Traditional CSMA protocols would require the
relays to sense that collision, back-off and retry. In that way
collision probability could be further reduced, at the expense of
increased latency overhead for relay selection. We will analyze
the collision probability without any contention resolution
protocol and further improvements are left for future work.

We will provide an analytic way to calculate a close-form
expression of equation (\ref{eq:upper_bound}) for any kind of
wireless fading statistics. But before doing so, we can easily
show that this probability can be made arbitrary small, close to
zero.

If $T_b=\min\{T_j\}, j \in [1,M]$ and $Y_1 < Y_2 < \ldots <Y_M$
the ordered random variables $\{T_j\}$ with $T_b \equiv Y_1$, and
$Y_2$ the second minimum timer, then:

\vspace{-2em}

\begin{equation}\label{eq:ordered}
    Pr(any ~T_j < T_b+c~|~j \neq b) \equiv Pr(Y_2 < Y_1+c)
\end{equation}
From the last equation, we can see that this probability can be
made arbitrarily small by decreasing the parameter $c$. For short
range radios (on the order of 100 meters), this is primarily
equivalent to selecting radios with small switch times (from
receive to transmit mode) on the order of a few microseconds.

Given that $Y_j=\lambda / h_{(j)}$, $Y_1 < Y_2 < \ldots <Y_M$ is
equivalent to $1/h_{(1)}<1/h_{(2)}<\ldots <1/h_{(M)}$\footnote{The
parenthesized subscripts are due to ordering of the channel
gains.}, equation (\ref{eq:ordered}) is equivalent to

\vspace{-2em}

\begin{equation}\label{eq:collision_asymptotic}
    Pr(Y_2 < Y_1+c)=Pr(\frac{1}{h_{(2)}} < \frac{1}{h_{(1)}}+\frac{c}{\lambda})
\end{equation}
and $Y_1 < Y_2 < \ldots <Y_M \Leftrightarrow h_{(1)}> h_{(2)}
\ldots > h_{(M)}$ ($h, \lambda, c$ are positive numbers).

From the last equation (\ref{eq:collision_asymptotic}), it is
obvious that increasing $\lambda$ at each relay (in equation
(\ref{eq:timer})), reduces the probability of collision to zero
since equation (\ref{eq:collision_asymptotic}) goes to zero with
increasing $\lambda$.

In practice, $\lambda$ can not be made arbitrarily large, since it
also "regulates" the expected time, needed for the network to find
out the "best" relay. From equation (\ref{eq:timer}) and Jensen's
inequality we can see that
\begin{equation}\label{eq:expected_time}
    E[T_j]=E[\lambda / h_j]\geq \lambda / E[h_j]
\end{equation}
or in other words, the expected time needed for each relay to flag
its presence, is lower bounded by $\lambda$ times a constant.
Therefore, there is a tradeoff between probability of collision
and speed of relay selection. We need to have $\lambda$ as big as
possible to reduce collision probability and at the same time, as
small as possible, to quickly select the best relay, before the
channel changes again (i.e. within the coherence time of the
channel). For example, for a mobility of $0-3$ km/h, the maximum
Doppler shift is $f_m=2.5~Hz$ which is equivalent with a minimum
coherence time on the order of $200$ milliseconds. Any relay
selection should occur well before that time interval with a
reasonably small probability of error. From figure
\ref{fig_ricean}, we note that selecting $c/\lambda \approx 1/200$
will result in a collision probability less than $0.6\%$ for
policy I.  Typical switching times result in $c \approx 5 \mu s$.
This gives $\lambda \approx 1 ms$ which is two orders of magnitude
less than the coherence interval. More sophisticated radios with
$c \approx 1 \mu s$ will result in $\lambda \approx 200 \mu s$,
which is three orders of magnitude smaller than the coherence time
\footnote{Note that the expected value of the minimum of the set
of random variables(timers) is smaller than the average of those
random variables. So we expect the overhead to be much smaller
than the one calculated above}.

\subsection{Calculating $Pr(Y_2 < Y_1 +c)$}
In order to calculate the collision probability from
(\ref{eq:ordered}), we first need to calculate the joint
probability distribution of the minimum and second minimum of a
collection  of $M$ i.i.d\footnote{The choice of identically
distributed timer functions implicitly assumes that the relays are
distributed in the same geographical region and therefore have
similar distances towards source and destination. In that case,
randomization among the timers is provided only by
\textit{fading}. The cases where the relays are randomly
positioned and have in general different distances, is a scenario
where randomization is provided not only because of fading, but
also because of different moments. In such asymmetric cases the
collision probability is expected to decrease and a concrete
example is provided.} random variables, corresponding to the timer
functions of the $M$ relays. The following theorem provides this
joint distribution:
\begin{theorem}
\label{theorem_jointpdf} The joint probability density function of
the minimum and second minimum among $M \geq 2$ i.i.d. positive
random variables $T_1,~T_2, \dots, ~T_M$, each with probability
density function $f(t)\equiv \frac{dF(t)}{dt}$ and cumulative
distribution function $F(t)\equiv Pr(T\leq t)$, is given by the
following equation:
\begin{equation}
 \nonumber
 f_{Y_1, Y_2}(y_1,y_2)=
 \begin{cases}
  M~(M-1)~f(y_1)~f(y_2)~[1-F(y_2)]^{M-2} &  for ~0<y_1<y_2 \\
  0 & elsewhere.
   \end{cases}
\end{equation}
where $Y_1<Y_2<Y_3 \ldots <Y_M$ are the $M$ ordered random
variables $T_1,~T_2, \ldots, ~T_M$.
\end{theorem}

\begin{proof}
Please refer to appendix \ref{appendix1}.
\end{proof}

Using Theorem 1, we can show the following lemma that gives a
closed-form expression for the collision probability (equation
\ref{eq:ordered}):

\begin{lemma}
\label{theorem_Pcollision} Given $M \geq 2$ i.i.d. positive random
variables $T_1,~T_2, \dots, ~T_M$, each with probability density
function $f(x)$ and cumulative distribution function $F(x)$, and
$Y_1<Y_2<Y_3 \ldots <Y_M$ are the $M$ ordered random variables
$T_1,~T_2, \ldots, ~T_M$, then $Pr(Y_2<Y_1 + c)$, where $c>0$, is
given by the following equations:
\begin{equation}\label{eq:collision}
Pr(Y_2<Y_1 + c) = 1-I_c
\end{equation}
\begin{equation}
I_c = M~(M-1)~\int_c^{+\infty} f(y)~[1-F(y)]^{M-2}~F(y-c)~dy
\end{equation}
\end{lemma}

\begin{proof}\\
Please refer to appendix \ref{appendix1}.
\end{proof}

Notice that the statistics of each timer $T_i$ and the statistics
of the wireless channel are related according to equation
(\ref{eq:timer}). Therefore, the above formulation is applicable
to any kind of wireless channel distribution.

\subsection{Results}

In order to exploit theorem \ref{theorem_jointpdf} and lemma
\ref{theorem_Pcollision}, we first need to calculate the
probability distribution of $T_i$ for $i ~\in [1,M]$. From
equation (\ref{eq:timer}) it is easy to see that the cdf $F(t)$
and pdf $f(t)$ of $T_i$ are related to the respective
distributions of $h_i$ according to the following equations:
\begin{equation}\label{eq:cdf}
    F(t)\equiv ~CDF_{T_i}(t)=Pr\{T_i~\leq~t\}=1-CDF_{h_i}(\frac{\lambda}{t})
\end{equation}
\begin{equation}\label{eq:pdf}
    f(t)\equiv ~pdf_{T_i}(t)=\frac{d}{dt}F(t)=\frac{\lambda}{t^2}~pdf_{h_i}(\frac{\lambda}{t})
\end{equation}

After calculating equations (\ref{eq:cdf}), (\ref{eq:pdf}), and
for a given $c$ calculated from (\ref{eq:c_nohidden}) or
(\ref{eq:c_hidden}), and a specific $\lambda$, we can calculate
probability of collision using equation (\ref{eq:collision}).

Before proceeding to special cases, we need to observe that for a
given distribution of the wireless channel, collision performance
depends on the ratio $c/\lambda$, as can be seen from equation
(\ref{eq:collision_asymptotic}), discussed earlier.

\subsubsection{Rayleigh Fading} Assuming $|a_{si}|, ~|a_{id}|$,
for any $i \in [1,M]$, are independent (but not identically
distributed) Rayleigh random variables,  then  $|a_{si}|^2,
~|a_{id}|^2$ are independent, exponential random variables, with
parameters $\beta_1, \beta_2$ respectively
($E[|a_{si}|^2]=1/\beta_1, ~E[|a_{id}|^2]=1/\beta_2$).

Using the fact that the minimum of two independent exponential
r.v.'s with parameters $\beta_1, \beta_2$, is again an exponential
r.v with parameter $\beta_1 +\beta_2$, we can calculate the
distributions for $h_i$ under policy I (equation
\ref{eq:policy1}). For policy II (equation \ref{eq:policy2}), the
distributions of the harmonic mean, have been calculated
analytically in \cite{Hasna03}. Equations (\ref{eq:cdf}) and
(\ref{eq:pdf}) become:

under policy I:
\begin{eqnarray}
  F(t) &=& e^{-~(\beta_1 + \beta_2) ~\lambda /t} \label{eq:f_policy1}\\
  f(t) &=& \frac{~\lambda~(\beta_1 + \beta_2)}{t^2}~ e^{-(\beta_1 + \beta_2) ~\lambda /t}
\end{eqnarray}

under policy II:
\begin{eqnarray}
  F(t) &=& \frac{\lambda ~\sqrt{\beta_1 ~ \beta_2}}{t}~ e^{-\lambda ~(\beta_1+\beta_2) /(2t)} ~K_1({\frac{\lambda ~\sqrt{\beta_1 \beta_2}}{t}}) \\
  f(t) &=&  ~\frac{\lambda^2}{2~t^3} ~\beta_1~\beta_2 ~e^{-\lambda ~(\beta_1 + \beta_2) /(2t)}
  ~[\frac{\beta_1+\beta_2}{\sqrt{\beta_1~\beta_2}}K_1({\frac{\lambda ~\sqrt{\beta_1~\beta_2}}{t}}) + 2~K_0({\frac{\lambda ~\sqrt{\beta_1~\beta_2}}{t}})]
  \label{eq:f_policy2}
\end{eqnarray}
where $K_i(x)$ is the modified Bessel function of the second kind
and order $i$.

\begin{figure}
\centering
\includegraphics[width=3.9in]{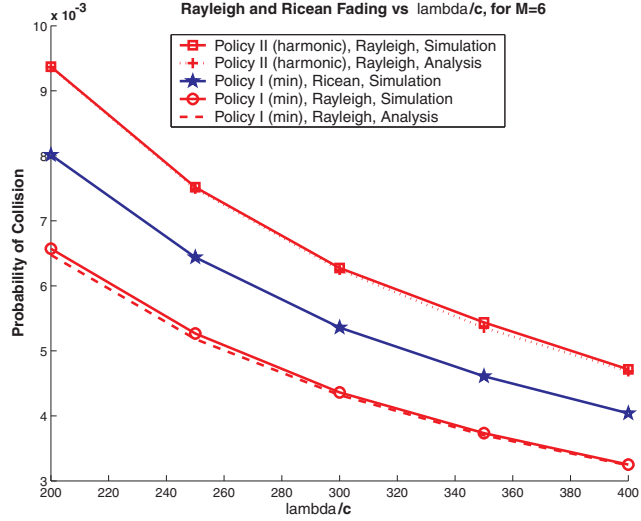}
\caption{Performance in Rayleigh and Ricean fading, for policy I
(min) and Policy II (harmonic mean), various values of ratio
$\lambda/c$ and $M=6$ relays, clustered at the same region. Notice
that collision probability drops well below $1\%$.}
\label{fig_ricean}
\end{figure}

Equation (\ref{eq:collision}) is calculated for the two policies,
for the symmetric case
($\beta_1=\beta_2=E[|a_{si}|^2]=E[|a_{id}|^2]=1$) of $M=6$ relays.
Monte-Carlo simulations are also performed under the same
assumptions. Results are plotted in fig. \ref{fig_ricean}, for
various ratios $\lambda/c$. We can see that Monte-carlo
simulations match the results provided by numerical calculation of
equation (\ref{eq:collision}) with the help of equations
(\ref{eq:f_policy1})-(\ref{eq:f_policy2}).

 Collision probability drops with increasing ratio of $\lambda/c$
 as expected. Policy I ("the minimum"), performs significantly
 better than Policy II ("the harmonic mean") and that can be
 attributed to the fact that the harmonic mean smooths the two
 path SNRs (between source-relay and relay-destination) compared
 to the minimum function. Therefore, the effect of
 randomization due to fading among the relay timers, becomes less
 prominent under Policy II. The probability can be kept well below $1\%$,
 for ratio $\lambda/c$ above $200$.

\subsubsection{Ricean Fading} It was interesting to examine the
performance of opportunistic relay selection, in the case of
Ricean fading, when there is a dominating communication path
between any two communicating points, in addition to many
reflecting paths and compare it to Rayleigh fading, where there is
a large number of equal power, independent paths.

Keeping the average value of any channel coefficient the same
($E[|a|^2]=1$) and assuming a single dominating path and a sum of
reflecting paths (both terms with equal total power), we plotted
the performance of the scheme when policy I was used, using
Monte-Carlo simulations (fig. \ref{fig_ricean}). We can see that
in the Ricean case, the collision probability slightly increases,
since now, the realizations of the wireless paths along different
relays are clustered around the dominating path and vary less,
compared to Rayleigh fading. Policy II performs slightly worse,
for the same reasons it performed slightly worse in the Rayleigh
fading case and the results have been omitted.

In either cases of wireless fading (Rayleigh or Ricean), the
scheme performs reasonably well.

\subsubsection{Different topologies}
\begin{figure}
\centering
  \hfill
  \includegraphics[width=2.2in]{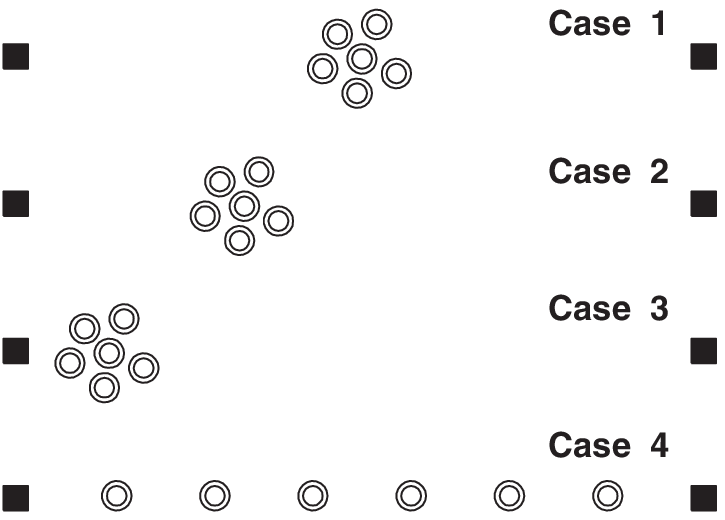}
  \hfill
  \includegraphics[width=3.9in]{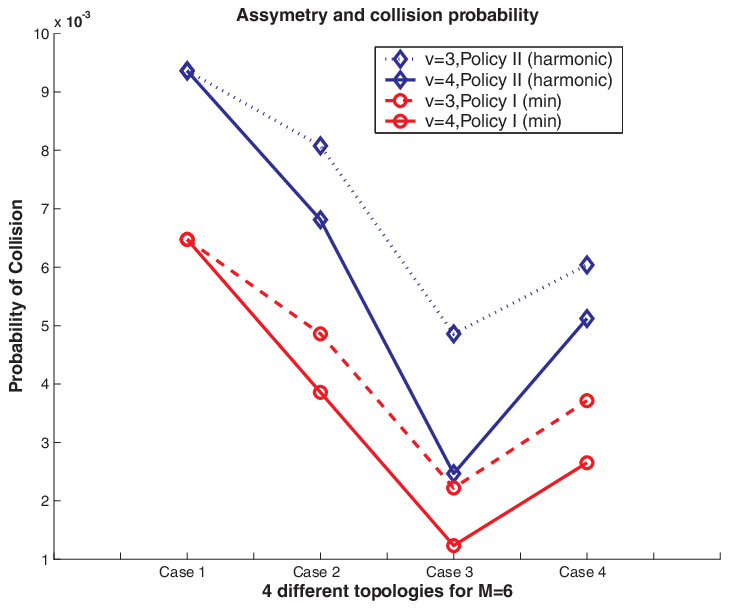}
  \hfill
  \caption{Unequal expected values (moments) among the two path SNRs or
  among the relays, reduce collision probability. M=6 and
  $c/\lambda = 1/200$ for the four different topologies considered.}
  \label{fig:topology}
\end{figure}

For the case of all relays not equidistant to source or
destination, we expect the collision probability to drop, compared
to the equidistant case, since the asymmetry between the two links
(from source to relay and from relay to destination) or the
asymmetry between the expected SNRs among the relays,   will
increase the variance of the timer function, compared to the
equidistant case. To demonstrate that, we studied three cases,
where $M=6$ relays are clustered half-way ($d/2$), closer to
transmitter ($d/3$) or even closer to transmitter ($d/10$) (case
1,2,3 respectively in fig. \ref{fig:topology} and $d$ is the
distance between source and destination) and one case where the
relays form an equidistant line network between source and
destination (case 4 in fig. \ref{fig:topology}).

Assuming Rayleigh fading, $c/\lambda=1/200$ and expected path
strength as a non-linear, decreasing function of distance
($E[|a_{ij}|^2]=1/\beta_{ij} \propto (1/d_{ij})^v$), we calculated
the collision probability for $M=6$ relays, using expressions
(\ref{eq:f_policy1})-(\ref{eq:f_policy2}) into
(\ref{eq:collision}) for cases 1, 2, 3 while for case 4 we used
Monte-Carlo simulation: in case 1, $\beta_1=\beta_2=1$, in case 2,
$\beta_1=(2/3)^v, \beta_2=(4/3)^v$ and in case 3,
$\beta_1=(1/5)^v, \beta_2=(9/5)^v$. For case 4, $\beta_1=(2/7)^v,
\beta_2=(12/7)^v$ for the closest terminal to source,
$\beta_1=(4/7)^v, \beta_2=(10/7)^v$ for the second closest
terminal to source, $\beta_1=(6/7)^v, \beta_2=(8/7)^v$ for the
third closest to source terminal. Due to symmetry, the expected
power and corresponding $\beta$ factors of the paths, for the
third closer to destination, second closer to destination and
closest terminal to destination, are the same with the ones
described before (third closer terminal to source, second closer
terminal to source and closest to source terminal respectively),
with $\beta_1$ and $\beta_2$ interchanged.

From fig. \ref{fig:topology}, we can see that the collision
probability of asymmetric cases 2, 3 and 4 is strictly smaller
compared to the symmetric case 1. Policy I performs better than
Policy II and collision probability decreases for increasing
factor $v$ ($v=3,4$ were tested). This observation agrees with
intuition that suggests that different moments for the path
strengths among the relays, increase the randomness of the
expiration times among the relays and therefore decrease the
probability of having two or more timers expire within the same
time interval.
%

We note that the source can also participate in the process of
deciding the best relay. In this special case, where the source
can receive the CTS message, it could have its own timer start
from a value depending upon the instantaneous $|a_{sd}|^2$. This
will be important if the source is not aware whether there are any
relays in the vicinity that could potentially cooperate.

The proposed method as described above, involving instantaneous
SNRs as a starting point for each relay's timer and using
\textit{time} (corresponding to an assessment of how good is a
particular path within the coherence time of the channel) to
select \textit{space} (the best available path towards
destination) in a distributed fashion, is novel and has not been
proposed before, to the best extent of our knowledge.

\section{Simplifying Cooperative Diversity through Opportunistic Relaying}
\label{Simplifying} We now consider the impact of opportunistic
relaying on the cooperative diversity scenario. The main result of
this section is that opportunistic relaying can be used to
simplify a number of cooperative diversity protocols involving
multiple relays. In particular we focus on the cooperative
diversity protocol in \cite{LanemanWornell03} which requires the
relays to use a space-time code while simultaneously transmitting
towards the destination. We show that this protocol can be
simplified considerably by simply selecting the best relay in the
second stage. Perhaps surprisingly, this simplified protocol
achieves the same diversity multiplexing tradeoff achieved in
\cite{LanemanWornell03}. Furthermore, it does not matter whether
the relay implements an amplify and forward or a decode and
forward protocol in terms of the diversity-multiplexing tradeoff.
We also note that opportunistic relaying can be used to simplify
the non-orthogonal relaying protocols proposed in
\cite{AzarianElGamal04}. However the detailed performance analysis
is left for future work.

\subsection{Channel Model}
We consider an i.i.d\ slow Rayleigh fading channel model following
\cite{LanemanTseWornell04}. A half duplex constraint is imposed
across each relay node, i.e. it cannot transmit and listen
simultaneously. We assume that the nodes (transmitter and relays)
do not exploit the knowledge of the channel at the physical layer.
Note that in the process of discovering the best relay described
in the previous section the nodes do learn about their channel
gains to the destination. However, we assume that this knowledge
of channel gain is limited to the network layer protocol. The
knowledge of channel gain is not exploited at the physical layer
in order to adjust the code rate based on instantaneous channel
measurements. In practice, the hardware at the physical layer
could be quite constrained to allow for this flexibility to change
the rate on the fly. It could also be that the transmitter is
operating at the minimum transmission rate allowed by the radio
hardware. Throughout this section, we assume that the channel
knowledge is not exploited at the physical layer at either the
transmitter or the relays.

If the discrete time received signal at the destination and the
relay node are denoted by $Y[n]$ and $Y_1[n]$ respectively:

\vspace{-2em}

\begin{eqnarray}
\label{eq_ch_mod_rx} Y[n] &=& \aasd X[n]   + Z[n], ~~~ n=1,2\ldots \frac{T}{2} ~~~ \text{(source~transmits~destination~receives)} \\
                    Y[n] &=& \aard X_1[n] + Z[n], ~~~ n = \frac{T}{2},\frac{T}{2}+1\ldots,T ~~ \text{(best~relay~transmits~destination~receives)} \\
\label{eq_ch_mod_rel} Y_1[n] &=& \aasr X[n] + Z_1[n] ~~~
n=1,2\ldots \frac{T}{2}
~~~\text{(source~transmits~best~relay~receives)}
\end{eqnarray}

Here $\aasd,\aard,\aasr$ are the respective channel gains from the
source to destination, best relay to destination and source to the
best relay respectively.  The channel gains between any two pair
of nodes are i.i.d\ $\cN(0,1)$\footnote{The channel gains from the
best relay to destination and source to best relay are not
$\cN(0,1)$. See Lemma \ref{lem:best-relay-gains} in the
Appendix.}. The noise $Z[n]$ and $Z_1[n]$ at the destination and
relay are both assumed to be i.i.d\ circularly symmetric complex
Gaussian $\cN(0,\sigma^2)$. $X[n]$ and $X_1[n]$ are the
transmitted symbols at the transmitter and relay respectively. $T$
denotes the duration of time-slots reserved for each message and
we assume that the source and the relay each transmit orthogonally
on half of the time-slots. We impose a power constraint at both
the source and the relay: $E[|X[n]|^2] \leq P$ and $E[|X_1[n]|^2]
\leq P$. For simplicity, we assume that both the source and the
relay to have the same power constraint. We will define $\rho
\stackrel{\Delta}{=} P/\sigma^2$ to be the effective signal to
noise ratio (SNR). This setting can be easily generalized when the
power at the source and relays is different.

The following notation is necessary in the subsequent sections of
the paper. This notation is along the lines of
\cite{AzarianElGamal04} and simplifies the exposition.

\begin{definition}
\label{def:exp-function} A function $f(\rho)$ is said to be
exponentially equal to $b$, denoted by $f(\rho) \doteq \rho^b$, if
\begin{equation}
\lim_{\rho\rightarrow\infty}\frac{\log f(\rho)}{\log \rho} = b.
\end{equation}

We can define the relation $\stackrel{.}{\leq}$ in a similar
fashion.

\end{definition}

\begin{definition}
\label{def:exp-order}
 The exponential order of a random variable
$X$ with a non-negative support is given by,
\begin{equation}
 V = -\lim_{\rho\rightarrow \infty} \frac{\log X}{\log\rho}.
\end{equation}

\end{definition}

The exponential order greatly simplifies the analysis of outage
events while deriving the diversity multiplexing tradeoff. Some
properties of the exponential order are derived in Appendix
\ref{appendix2}, lemma \ref{lem:exp-order-max}.

\begin{definition}({\bf Diversity-Multiplexing Tradeoff}) We use the
definition given in \cite{ZhengTse03}. Consider a family of codes
$C_\rho$ operating at SNR $\rho$ and having rates $R(\rho)$ bits
per channel use. If $P_e(R)$ is the outage probability (see
\cite{Telatar99}) of the channel for rate $R$, then the
multiplexing gain $r$ and diversity order $d$ are defined
as\footnote{We will assume that the block length of the code is
large enough, so that the detection error is arbitrarily small and
the main error event is due to outage.}

\begin{minipage}{5cm}
\begin{equation*}
r \stackrel{\Delta}{=}
\lim_{\rho\rightarrow\infty}\frac{R(\rho)}{\log \rho}
\end{equation*}
\end{minipage}
\begin{minipage}{5.5 cm}
\begin{equation}
d \stackrel{\Delta}{=} -\lim_{\rho\rightarrow\infty}\frac{\log
P_e(R)}{\log \rho} \label{eq:div-gain}\\\\
\end{equation}
\end{minipage}
\end{definition}

What remains to be specified is a policy for selecting the best
relay. We essentially use the policy 1 (equation
(\ref{eq:policy1})) in the previous section.

\begin{myrule} Among all the available relays, denote the relay with
the largest value of $\min\{\asr^2,\ard2\}$ as the best relay.
\label{lb:rule1}
\end{myrule}

To justify this choice, we note from fig. \ref{fig_mac} that the
performance of policy I is slightly better than policy II.
Furthermore, we will see in this section that this choice is
optimum in that it enables opportunistic relaying to achieve the
same diversity multiplexing tradeoff of more complex orthogonal
relaying schemes in \cite{LanemanWornell03}. We next discuss the
performance of the amplify and forward and decode and forward
protocols.

\subsection{Digital Relaying - Decode and Forward Protocol}
We will first study the case where the intermediate relays have
the ability to decode the received signal, re-encode and transmit
it to the destination. We will study the protocol proposed in
\cite{LanemanWornell03} and show that it can be considerably
simplified through opportunistic relaying.

The decode and forward algorithm considered in
\cite{LanemanWornell03} is briefly summarized as follows. In the
first half time-slots the source transmits and all the relays and
receiver nodes listen to this transmission. Thereafter {\em all}
the relays that are successful in decoding the message, re-encode
the message using a distributed space-time protocol and
collaboratively transmit it to the destination. The destination
decodes the message at the end of the second time-slot. Note that
the source does not transmit in the second half time-slots. The
main result for the decode and forward protocol is given in the
following theorem                                       :

\begin{theorem}[\cite{LanemanWornell03}] The achievable diversity
multiplexing tradeoff for the decode and forward strategy with $M$
intermediate relay nodes is given by $d(r) = (M+1)(1-2r)$ for $r
\in (0,0.5)$. \label{thm:BasicDF}
\end{theorem}


The following Theorem shows that opportunistic relaying achieves
the same diversity-multiplexing tradeoff if the best relay
selected according to policy \ref{lb:rule1}.

\begin{theorem} Under opportunistic relaying, the decode and
forward protocol with $M$ intermediate relays achieves the same
diversity multiplexing tradeoff stated in Theorem
\ref{thm:BasicDF}. \label{thm:BasicDF-opp}
\end{theorem}

\begin{proof}
We follow along the lines of \cite{LanemanWornell03}. Let $\E$
denote the event that the relay is successful in decoding the
message at the end of the first half of transmission and
$\bar{\E}$ denote the event that the relay is not successful in
decoding the message. Event $\bar{\E}$ happens when the mutual
information between source and best relay drops below the code
rate. Suppose that we select a code with rate $R = r\log\rho$ and
let $I(X;Y)$ denote the mutual information between the source and
the destination. The probability of outage is given by
\begin{eqnarray*}
P_e &=& \Pr(I(X;Y)\leq r\log\rho|\E)\Pr(\E) + \Pr(I(X;Y)\leq
r\log\rho|\bar{\E})\Pr(\bar{\E}) \\
&=& \Pr\left(\frac{1}{2}\log(1 + \rho(\asd^2 + \ard^2))\leq
r\log \rho\right)\Pr(\E) + \\
&& \Pr\left(\frac{1}{2}\log(1 + \rho\asd^2)\leq r\log
\rho\right)\Pr(\bar{\E}) 
\end{eqnarray*}
\begin{eqnarray*}
&\leq& \Pr\left(\frac{1}{2}\log(1 + \rho(\asd^2 + \ard^2))\leq
r\log \rho\right)+ \\
&& \Pr\left(\frac{1}{2}\log(1 + \rho\asd^2)\leq r\log
\rho\right)\Pr\left(\frac{1}{2}\log(1 + \rho\asr^2)\leq r\log
\rho\right) \\
&\leq& \Pr\left(\asd^2 + \ard^2\leq \rho^{2r-1} \right)+
\Pr\left(\asd^2 \leq \rho^{2r-1}\right)\Pr\left(\asr^2\leq
\rho^{2r-1}\right) \\
&\leq& \Pr\left(\asd^2 \leq \rho^{2r-1} \right)\Pr\left(\ard^2
\leq \rho^{2r-1} \right) + \Pr\left(\asd^2 \leq
\rho^{2r-1}\right)\Pr\left(\asr^2\leq
\rho^{2r-1}\right) \\
&\dotleq& \rho^{2r-1}\rho^{M(2r-1)} + \rho^{2r-1}\rho^{M(2r-1)}
\doteq \rho^{(M+1)(2r-1)}
\end{eqnarray*}

In the last step we have used claim 2 of Lemma
\ref{lem:best-relay-gains} in the appendix with $m=M$.

\end{proof}

We next study the performance under analog relaying and then
mention several remarks.

\subsection{Analog relaying - Basic Amplify and Forward}

We will now consider the case where the intermediate relays are
not able to decode the message, but can only scale their received
transmission (due to the power constraint) and send it to the
destination.

The basic amplify and forward protocol was studied in
\cite{LanemanTseWornell04} for the case of a single relay. The
source broadcasts the message for first half time-slots. In the
second half time-slots the relay simply amplifies the signals it
received in the first half time-slots. Thus the destination
receives two copies of each symbol. One directly from the source
and the other via the relay. At the end of the transmission, the
destination then combines the two copies of each symbol through a
matched filter. Assuming i.i.d\ Gaussian codebook, the mutual
information between the source and the destination can be shown to
be \cite{LanemanTseWornell04},

\vspace{-2em}

\begin{eqnarray}
\label{eq:basic-AF} I(X;Y) &=& \frac{1}{2}\log\left(1 + \rho
\asd^2 + f(\rho \asr^2, \rho\ard^2) \right)\\
\label{eq:basic-fexp}f(a,b) &=& \frac{ab}{a+b+1}
\end{eqnarray}

The amplify and forward strategy does not generalize in the same
manner as the decode and forward strategy for the case of multiple
relays. We do not gain by having all the relay nodes amplify in
the second half of the time-slot. This is because at the
destination we do not receive a coherent summation of the channel
gains from the different receivers. If $\gamma_j$ is the scaling
constant of receiver $j$, then the received signal will be given
by $y[n] = \left(\sum_{j=1}^{M}\gamma_j
a_{rd}^{j}\right)x[n]+z[n].$ Since this is simply a linear
summation of Gaussian random variables, we do not see the
diversity gain from the relays. A possible alternative is to have
the $M$ relays amplify in a round-robin fashion. Each relay
transmits only one out of every $M$ symbols in a round robin
fashion. This strategy has been proposed in
\cite{LanemanWornell03}, but the achievable diversity-multiplexing
tradeoff is not analyzed.

Opportunistic relaying on the other hand provides another possible
solution to analog relaying. Only the best relay (according to
policy \ref{lb:rule1}) is selected for transmission. The following
theorem shows that opportunistic relaying achieves the same
diversity multiplexing tradeoff as that achieved by the (more
complicated) decode and forward scheme.

\begin{theorem}Opportunistic amplify and forward achieves the same
diversity multiplexing tradeoff stated in Theorem
\ref{thm:BasicDF}. \label{thm:OppAF}
\end{theorem}

\begin{proof}
We begin with the expression for mutual information between the
source and destination  (\ref{eq:basic-AF}). An outage occurs if
this mutual information is less than the code rate $r\log\rho$.
Thus we have that
\begin{eqnarray*}
P_e &=& \Pr\left(I(X;Y) \leq r\log\rho\right)\\
&=& \Pr\left(\log(1+\rho\asd^2+f(\rho\asr^2,\rho\ard^2)\leq
2r\log\rho\right)\\
&\leq&\Pr\left(\asd^2\leq \rho^{2r-1},
f(\rho\asr^2,\rho\ard^2)\leq \rho^{2r}\right)\\
&\stackrel{(a)}{\leq}&\Pr\left(\asd^2\leq \rho^{2r-1},
\min{(\asr^2,\ard^2)}\leq \rho^{2r-1} + \rho^{r-1}\sqrt{1+\rho^{2r}}\right)\\
&\stackrel{(b)}{\doteq}&\rho^{2r-1}\rho^{M(2r-1)} =
\rho^{(M+1)(2r-1)}
\end{eqnarray*}
Here (a) follows from Lemma \ref{lem:fminrel} and (b) follows from
Lemma \ref{lem:best-relay-gains}, claim 1 in appendix
\ref{appendix2} and the fact that $\rho^{r-1}\sqrt{1+\rho^{2r}}
\rightarrow \rho^{2r-1}$ as $\rho \rightarrow \infty$.
\end{proof}

\begin{figure}
  \centering
  \includegraphics[width=4.0in]{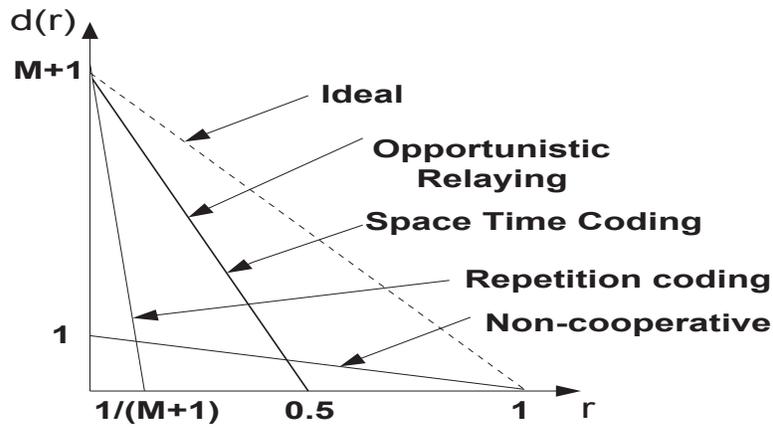}\\
  \caption{The diversity-multiplexing of opportunistic relaying is exactly the same with that of
  more complex space-time coded protocols.}\label{fig:tradeoff}
\end{figure}

\subsection{Discussion}

\subsubsection{Space-time Coding vs. Relaying Solutions}

The (conventional) cooperative diversity setup (e.g.
\cite{LanemanWornell03}) assumes that the cooperating relays use a
distributed space-time code to achieve the diversity multiplexing
tradeoff in Theorem \ref{thm:BasicDF}. Development of practical
space-time codes is an active area of research. Recently there has
been considerable progress towards developing practical codes that
achieve the diversity multiplexing tradeoff over MIMO channels. In
particular it is known that random lattice based codes (LAST) can
achieve the entire diversity multiplexing tradeoff over MIMO
channels \cite{GamalCaireDamen}. Moreover it is noted in
\cite{Nabar} that under certain conditions the analytical
criterion such as rank and determinant criterion for MIMO links
also carry over to cooperative diversity systems \footnote{However
it is assumed that the destination knows the channel gain between
source and relay for Amplify and Forward.}. However some practical
challenges will have to be addressed to use these codes in the
distributed antenna setting: (a) The codes for MIMO channels
assume a fixed number of transmit and receive antennas. In
cooperative diversity, the number of antennas depends on which
relays are successful in decoding and hence is a variable
quantity. (b) The destination must be informed either explicitly
or implicitly which relays are transmitting.


Opportunistic relaying  provides an alternative solution to space
time codes for cooperative diversity by using a clever relaying
protocol. The result of Theorem \ref{thm:OppAF} suggests that
there is no loss in diversity multiplexing
tradeoff\footnote{compared to the orthogonal transmission
protocols in \cite{LanemanWornell03}} if a simple analog relaying
based scheme is used in conjunction with opportunistic relaying.
Even if the intermediate relays are digital, a very simple decode
and forward scheme that does eliminates the need for space-time
codes can be implemented. The relay listens and decodes the
message in the first half of the time-slots and repeats the source
transmission in the second half of the time-slots when the source
is not transmitting. The receiver simply does a maximal ratio
combining of the source and relay transmissions and attempts to
decode the message. Theorem \ref{thm:BasicDF-opp} asserts that
once again the combination of this simple physical layer scheme
and the smart choice of the relay is essentially optimum.

The diversity-multiplexing tradeoff is plotted in fig.
\ref{fig:tradeoff}. Even though a single terminal with the ``best"
end-to-end channel conditions relays the information, the
diversity order in the high SNR regime is on the order of the
number $M+1$ of all participating terminals. Moreover, the
tradeoff is exactly the same with that when space-time coding
across $M$ relays is used.

\subsubsection{Non-orthogonal Cooperative Diversity Schemes}

The focus in this paper was on the multiple relay cooperative
diversity protocols proposed in \cite{LanemanWornell03}, since
they require that the transmitter and relay operate in orthogonal
time-slots in addition to the half duplex constraints.  The
orthogonality assumption was amenable to practical implementation
\cite{Bletsas05_thesis}, since the decoder is extremely simple.
More recently, a new class of protocols that relax the assumption
that the transmitter and relay operate in orthogonal time-slots,
(but still assume the half duplex constraint) have been proposed
in \cite{AzarianElGamal04}. These protocols have a superior
performance compared to \cite{LanemanWornell03}, albeit at the
cost of higher complexity both at the decoder and network layer.
Opportunistic relaying could be naturally used to simplify those
protocols \footnote{An Alamouti\cite{Alamouti} type code could be
used if the relay and source are simultaneously transmitting.} and
details of such simplifications and its performance are underway.

%
%

\subsubsection{Impact of Topology} The analysis in for diversity-multiplexing tradeoff was presented assuming that average channel
gains between each pair of nodes is unity. In other words the
impact of topology was not considered. We observe that the effect
of topology can be included in the analysis using techniques used
in \cite{LanemanTseWornell04}. In the high SNR regime, we expect
fixed multiplicative factors of path loss to contribute little in
affecting the diversity-multiplexing tradeoff. However topology is
certainly important for finite SNR case as observed in
\cite{KramerGastparGupta04}.

\section{Conclusion}
\label{conclusion} We proposed Opportunistic Relaying as a
practical scheme for cooperative diversity. The scheme relies on
distributed path selection considering instantaneous end-to-end
wireless channel conditions, facilitates coordination among the
cooperating terminals with minimum overhead and could simplify the
physical layer in communicating transceivers by eliminating the
requirement of space time codes.

We presented a method to calculate the performance of the relay
selection algorithm, for any kind of wireless fading model and
showed that successful relay selection could be engineered with
reasonable performance. Specific examples for Rayleigh and Ricean
fading were given.

We treated Opportunistic Relaying as a distributed virtual antenna
array system and analyzed its diversity-multiplexing tradeoff,
revealing NO performance loss when compared with complex
space-time coding protocols in the field.

The approach presented in this work explicitly addresses
coordination among the cooperating terminals and has similarities
with a Medium Access Protocol (MAC) since it directs \textit{when}
a specific node to relay. The algorithm has also similarities with
a Routing Protocol since it coordinates \textit{which} node to
relay (or not) received information among a collection of
candidates. Devising wireless systems that dynamically adapt to
the wireless channel conditions without external means (for
example GPS receivers), in a distributed manner, similarly to the
ideas presented in this work, is an important and fruitful area
for future research.

The simplicity of the technique, allows for immediate
implementation in existing radio hardware and its adoption could
provide for improved flexibility, reliability and efficiency in
future 4G wireless systems.


%
%
\appendices
\section{Probabilistic Analysis of Successful Path Selection}
\label{appendix1}

\textit{Theorem} \ref{theorem_jointpdf} The joint probability
density function of the minimum and second minimum among $M$
i.i.d. positive random variables $X_1,~X_2, \dots, ~X_M$, each
with probability density function $f(x)$ and cumulative
distribution function $F(x)$, is given by the following equation:
\begin{equation}
 \nonumber
 f_{Y_1, Y_2}(y_1,y_2)=
 \begin{cases}
  M~(M-1)~f(y_1)~f(y_2)~[1-F(y_2)]^{M-2} &  for ~0<y_1<y_2 \\
  0 & elsewhere.
   \end{cases}
\end{equation}
where $Y_1<Y_2<Y_3 \ldots <Y_M$ are the $M$ ordered random
variables $X_1,~X_2, \ldots, ~X_M$.

\begin{proof}
$ f_{Y_1,Y_2}(y_1,y_2)~dy_1~dy_2 = Pr(Y_1~\in~dy_1,~Y_2~\in dy_2)
=
\\
Pr(one~X_i~in~dy_1,~one ~X_j~in ~dy_2 ~(with ~y_2>y_1 ~and ~i \neq
j), ~and ~all ~the ~rest ~X_i's ~greater ~than ~y_2)= \\
=~2~\binom{M}{2}~Pr(~X_1 \in dy_1, ~X_2 \in dy_2~(y_2>y_1),
~X_i>y_2, ~i \in ~[3,M])~=  \\
=~2~\binom{M}{2} f(y_1)~dy_1~f(y_2)~dy_2~[1-F(y_2)]^{M-2}= \\
=~M~(M-1)~f(y_1)~f(y_2)~[1-F(y_2)]^{M-2}~dy_1~dy_2, ~for
~0<y_1<y_2. $

The third equality is true since there are $\binom{M}{2}$ pairs in
a set of M i.i.d. random variables. The factor 2 comes from the
fact that ordering in each pair matters, hence we have a total
number of $2~\binom{M}{2}$ cases, with the same probability,
assuming identically distributed random variables. That concludes
the proof.
\end{proof}

Using Theorem 1, we can prove the following lemma:

\begin{figure}
\centering
\includegraphics[width=2.7in]{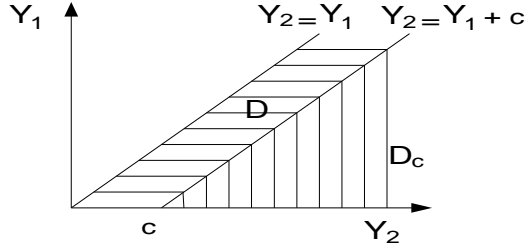}
\caption{Regions of integration of $f_{Y_1,Y_2}(y_1,y_2)$, for
$Y_1 < Y_2$ needed in Lemma I for calculation of $Pr(Y_2<Y_1+c), ~
c>0$.} \label{fig_region}
\end{figure}

\textit{Lemma} \ref{theorem_Pcollision} Given $M$ i.i.d. positive
random variables $X_1,~X_2, \dots, ~X_M$, each with probability
density function $f(x)$ and cumulative distribution function
$F(x)$, and $Y_1<Y_2<Y_3 \ldots <Y_M$ the $M$ ordered random
variables $X_1,~X_2, \ldots, ~X_M$, then $Pr(Y_2<Y_1 + c)$, where
$c>0$, is given by the following equations:
\begin{equation}\label{eq:collision2}
Pr(Y_2<Y_1 + c) = 1-I_c
\end{equation}
\begin{equation}
I_c = M~(M-1)~\int_c^{+\infty} f(y)~[1-F(y)]^{M-2}~F(y-c)~dy
\end{equation}

\begin{proof}
The joint pdf $f_{Y_1,Y_2}(y_1,y_2)$ integrates to $1$ in the
region $D \cup D_c$, as it can be seen in fig. \ref{fig_region}.
Therefore:
\begin{eqnarray}
 \nonumber
  Pr(~Y_2<Y_1+c) &=& \int \int_{D}~f_{Y_1,Y_2}(y_1,y_2)~dy1~dy2 \\ \nonumber
   &=& 1- \int \int_{D_c}~f_{Y_1,Y_2}(y_1,y_2)~dy1~dy2\\ \nonumber
   &=& 1-I_c \nonumber
\end{eqnarray}

Again from fig. \ref{fig_region}, $I_c$ can easily be calculated:
\begin{equation}
\nonumber I_c= \scriptstyle ~ M(M-1)~ \displaystyle
\int_{y_2=c}^{+\infty}f(y_2)~[1-F(y_2)]^{M-2}~\int_{0}^{y_2-c}f(y_1)~dy_1~dy_2
\end{equation}
\begin{equation}
=\scriptstyle ~M(M-1) \displaystyle
~\int_{y_2=c}^{+\infty}f(y_2)~[1-F(y_2)]^{M-2}~F(y_2-c)~dy_2
\end{equation}

The last equation concludes the proof.
\end{proof}


\section{Diversity-Multiplexing Tradeoff Analysis}
\label{appendix2}

We repeat \textit{Definition 1} and \textit{Definition 2} in this
section for completeness. The relevant lemmas follow.

\textit{Definition 1:} A function $f(\rho)$ is said to be
exponentially equal to $b$, denoted by $f(\rho) \doteq \rho^b$, if
\begin{equation}
\lim_{\rho\rightarrow\infty}\frac{\log f(\rho)}{\log \rho} = b.
\end{equation}

We can define the relation $\stackrel{.}{\leq}$ in a similar
fashion.
\textit{Definition 2:} The exponential order of a random
variable $X$ with a non-negative support is given by,
\begin{equation}
 V = -\lim_{\rho\rightarrow \infty} \frac{\log X}{\log\rho}.
\end{equation}
\begin{lemma}
Suppose $X_1,X_2,\ldots,X_m$ are $m$ i.i.d exponential random
variables with parameter $\lambda$ (mean $1/\lambda)$, and $X =
\max\{X_1,X_2,\ldots X_m\}$. If $V$ is the exponential order of
$X$ then the density function of $V$ is given by
\begin{equation}
\label{eq_exp_order_density} f_V(v) \doteq
\begin{cases}
\rho^{-mv} & v \geq 0 \\
0 & v < 0
\end{cases}
\end{equation}
and
\begin{equation}\label{eq:exp_order}
\Pr(X \leq \rho^{-v} ) \doteq \rho^{-mv}
\end{equation}

 \label{lem:exp-order-max}
\end{lemma}
\begin{proof}
Define,  $$V_\rho = -\frac{\log X}{\log\rho}.$$ Thus $V_\rho$ is
obtained from definition \ref{def:exp-order}, without the limit of
$\rho \rightarrow \infty$.
\begin{eqnarray*}
 \Pr(V_\rho\geq v) &=& \Pr(X \leq \rho^{-v} ) \\
    &=& \Pr(X_1 \leq \rho^{-v}, X_2 \leq \rho^{-v},\ldots X_m\leq
    \rho^{-v}) \\
    &=& \prod_{i=1}^m \Pr(X_i \leq \rho^{-v}) \\
    &=& \left(1-\exp(-\lambda\rho^{-v})\right)^m \\
    &=& \left(\lambda\rho^{-v}+ \sum_{j=2}^\infty
    \frac{(-\lambda)^j}{j!}~\rho^{-jv}\right)^m
\end{eqnarray*}

Note that $\Pr(V_\rho \geq v) \approx \rho^{-mv}$. Differentiating
with respect to $v$ and then taking the limit $\rho \rightarrow
\infty$, we recover (\ref{eq_exp_order_density}).
\end{proof}

From the above it can be seen that for the simple case of a single
exponential random variable ($m=1$), $\Pr(X \leq \rho^{-v}
)=\Pr(V_\rho\geq v) \doteq \rho^{-v}$.

\begin{lemma}
For relays, $j = 1,2,\ldots,m$, let $a_{sj}$ and $a_{jd}$ denote
the channel gains from source to relay $j$ and relay $j$ to
destination. Suppose that $\aasr$ and $\aard$ denote the channel
gain of the source to the best relay and the best relay to the
destination, where the relay is chosen according to rule
\ref{lb:rule1}. i.e.
$$\min(\asr^2,\ard^2) = \max\{\min(|a_{s1}|^2,|a_{1d}|^2),\ldots,\min(|a_{sm}|^2,|a_{md}|^2)\}$$
Then,
\begin{enumerate}

\item  $\min(\asr^2,\ard^2)$ has an exponential order given by
(\ref{eq_exp_order_density}).

\item
\[
\Pr(\asr^2 \leq \rho^{-v}) = \Pr(\ard^2 \leq \rho^{-v})
\stackrel{.}{\leq}
\begin{cases}
\rho^{-mv} & {v \geq 0 } \\
1 & \text{otherwise}
\end{cases}
\]
\label{eq-exp-order-joint}

\end{enumerate}
\label{lem:best-relay-gains}
\end{lemma}

\begin{proof} Let us denote $X^{(j)}
\stackrel{\Delta}{=} \min(|a_{sj}|^2,|a_{jd}|^2)$.
Since each of the $X^{(j)}$ are exponential random variables with
parameter 2, claim 1 follows from Lemma \ref{lem:exp-order-max}.
Also since $\asd^2$ and $\ard^2$ cannot be less than
$\min(\asd^2,\ard^2)$ claim 2 follows immediately from claim 1.
\end{proof}

\begin{lemma}

With $f(\cdot,\cdot)$ defined by relation (\ref{eq:basic-fexp}),
we have that \\
$~~~~~~~~~~~~~~~~~~\Pr\left(f(\rho a, \rho b) \leq
\rho^{2r}\right) \leq \Pr\left(\min(a,b)\leq
\rho^{2r-1}+\rho^{r-1}\sqrt{1+\rho^{2r}}\right)$.
\label{lem:fminrel}
\end{lemma}

\begin{proof}
Without loss in generality, assume that $a\geq b$.
\begin{eqnarray*}
f(\rho a, \rho b) &=& \rho\frac{ab}{a+b+\frac{1}{\rho}}\\
&=& \rho b \left(\frac{a}{a+b+\frac{1}{\rho}}\right)\\
&\stackrel{(a)}{\geq}& \rho
b\left(\frac{b}{2b+\frac{1}{\rho}}\right)
\end{eqnarray*}
Here (a) follows since $\frac{a}{a+K}$ is an increasing function
in $a$, for $K>0$ and $a\geq b$.

Now we have that
\begin{eqnarray*}
\Pr\left(f(\rho a, \rho b\right) \leq \rho^{2r}) &\leq&
\Pr\left(\frac{b^2}{2b+\frac{1}{\rho}}\leq \rho^{2r-1}\right)\\
&=&\Pr\left(b^2 \leq 2\rho^{2r-1}b + \rho^{2r-2}\right) \\
&=&\Pr\left((b-\rho^{2r-1})2 \leq \rho^{4r-2}+\rho^{2r-2}\right)
\\
&\stackrel{(a)}{=}&\Pr\left(b \leq
\rho^{2r-1}+\rho^{r-1}\sqrt{1+\rho^{2r}}\right)
\end{eqnarray*}

Where (a) follows since $b \geq 0$ so that $\Pr(b < 0) = 0$.

\end{proof}

\end{document}